\documentstyle[epsfig,multicol,pre,aps]{revtex}
\begin{document}
\draft
\title{Nonlinear Dynamics of A Damped Magnetic Oscillator
}
\author{
        Sang-Yoon Kim
        \footnote{Electronic address: sykim@cc.kangwon.ac.kr}
       }
\address{
Department of Physics\\ Kangwon National University\\
Chunchon, Kangwon-Do 200-701, Korea
}
\maketitle

\begin{abstract}
We consider a damped magnetic oscillator, consisting of a permanent
magnet in a periodically oscillating magnetic field. A detailed
investigation of the dynamics of this dissipative magnetic system
is made by varying the field amplitude $A$. As $A$ is increased,
the damped magnetic oscillator, albeit simple looking, exhibits
rich dynamical behaviors such as symmetry-breaking pitchfork
bifurcations, period-doubling transitions to chaos,
symmetry-restoring attractor-merging crises, and saddle-node
bifurcations giving rise to new periodic attractors. Besides
these familiar behaviors, a cascade of ``resurrections''
(i.e., an infinite sequence of alternating restabilizations
and destabilizations) of the stationary points also occurs.
It is found that the stationary points restabilize (destabilize)
through alternating subcritical (supercritical) period-doubling
and pitchfork bifurcations. We also discuss the critical behaviors
in the period-doubling cascades.
\end{abstract}


%
% End of Abstract
%

\begin{multicols}{2}
\section{Introduction}
\label{sec:Int}

We consider a permanent magnet of dipole moment $m$ placed in a
periodically oscillating magnetic field. This magnetic
oscillator (MO) can be described by a second-order non-autonomous
ordinary differential equation
\cite{Croquette,Schmidt1,Briggs},
\begin{equation}
I {\ddot \theta} + b {\dot  \theta} + m B_{0} \cos \omega t
\sin \theta =0,
\label{eq:MO1}
\end{equation}
where the overdot denotes the differentiation with respect to time,
$\theta$ is the angle between the permanent magnet and the magnetic
field, $I$ is the moment of inertia about a rotation axis, $b$ is
the damping parameter, and $B_{0}$ and $\omega$ are the amplitude
and frequency of the periodically oscillating magnetic field,
respectively.

Making the normalization
$\omega t \rightarrow 2 \pi (t+{1 \over 2})$ and $ \theta \rightarrow
2 \pi x$, we obtain a dimensionless form of Eq.~(\ref{eq:MO1}),
\begin{equation}
 {\ddot x} + \Gamma {\dot  x} - A \cos 2 \pi t \sin 2 \pi x =0,
\label{eq:MO2}
\end{equation}
where $x$ is a normalized angle with mod $1$,
$\Gamma={ {2 \pi b} / {I \omega}}$ and
$A = { {2 \pi m B_{0}} / {I \omega^2}}$. Note also that
Eq.~(\ref{eq:MO2}) describes the motion of a particle in a
standing wave field \cite{Bialek,Schmidt2,Escande}. For the
conservative case of $\Gamma=0$, the Hamiltonian system exhibits
period-doubling bifurcations and large-scale stochasticity as the
normalized field amplitude $A$ is increased, which have been found
both experimentally \cite{Croquette,Schmidt1,Briggs} and theoretically
\cite{Bialek,Schmidt2,Escande}. Here we are interested in the damped
case of $\Gamma \neq 0$ and make a detailed investigation of the
dynamical behaviors of the damped MO by varying the amplitude $A$.

This paper is organized as follows. We first discuss bifurcations
associated with stability of periodic orbits and Lyapunov exponents
in the damped MO in Sec.~\ref{sec:Bif}. With increasing $A$ up to
sufficently large values, dynamical behaviors of the damped MO are
then investigated in Sec.~\ref{sec:DB}. This very simple-looking
damped MO shows a richness in its dynamical behaviors. As $A$ is
increased, breakdown of symmetries via pitchfork bifurcations
\cite{Guckenheimer1}, period-doubling transitions to chaos
\cite{Feigenbaum}, restoration of symmetries via attractor-merging
crises \cite{Crisis}, birth of new periodic attractors through
saddle-node bifurcations \cite{Guckenheimer1}, and so on are
numerically found. In addition to these familiar behaviors, the
stationary points exhibit a cascade of ``resurrections'' \cite{Kim}
(i.e., they restabilize after their instability, destabilize again,
and so forth {\it ad infinitum}). It is found that the
restabilizations (destabilizations) occur via
alternating subcritical (supercritical) period-doubling and pitchfork
bifurcations. An infinite sequence of period-doubling
bifurcations, leading to chaos, also follows each destabilization
of the stationary points. In Sec.~\ref{sec:CSB}, we also study the
critical scaling behaviors in the period-doubling cascades.
It is found that the critical behaviors are the same as those for the
one-dimensional (1D) maps \cite{Feigenbaum}. Finally, a summary is
given in Sec.~\ref{sec:SUM}.

\section{Stability, Bifurcations and Lyapunov Exponents}
\label{sec:Bif}

In this section we first discuss stability of periodic orbits in the
damped MO, using the Floquet theory \cite{Guckenheimer2}. Bifurcations
associated with the stability and Lyapunov exponents are then
discussed.

The second-order ordinary differential equation (\ref{eq:MO2}) is
reduced to two first-order ordinary differential equations:
\begin{mathletters}
\begin{eqnarray}
{\dot x} &=& y,  \\
{\dot y} &=& -\Gamma y + A \cos 2 \pi t  \sin 2 \pi x.
\end{eqnarray}
\label{eq:MO3}
\end{mathletters}
These equations have two symmetries $S_1$ and $S_2$, because
the transformations
\begin{eqnarray}
S_1: && x \rightarrow x \pm {1 \over 2},~y \rightarrow y,
~t \rightarrow t \pm {1 \over 2}, \label{eq:S1} \\
S_2: && x \rightarrow -x,~y \rightarrow -y,~ t  \rightarrow t,
\label{eq:S2}
\end{eqnarray}
leave Eq.~(\ref{eq:MO3}) invariant.
The transformations in Eqs.~(\ref{eq:S1}) and (\ref{eq:S2}) are just
the shift in both $x$ and $t$ and the (space) inversion, respectively.
Hereafter, we will call $S_1$ and $S_2$ the shift and inversion
symmetries, respectively. If an orbit $z(t) [\equiv (x(y),y(t))]$
is invariant under $S_i$ $(i=1,2)$, it is called an
$S_i$-symmetric orbit. Otherwise, it is called an $S_i$-asymmetric
orbit and has its ``conjugate'' orbit $S_i z(t)$.

The phase space of the damped MO is three dimensional with the
coordinates $x$, $y,$ and $t$. Since the damped MO is periodic in $t$,
it is convenient to regard time as a circular coordinate (with mod
$1$) in the phase space. We then consider the surface of section, the
$x-y$ plane at interger times (i.e., $t=m$, $m$: integer). The
phase-space trajectiory intersects this plane in a sequence of points.
This sequence of points corresponds to a mapping on the plane. This
map plot of an initial point $z_0$ $[=(x_0,y_0)]$ can be conveniently
generated by sampling the orbit points $z_m$ at the discrete time
$t=m$. We call the transformation $z_m \rightarrow z_{m+1}$ the
Poincar\'{e} map and write $z_{m+1}= P (z_m)$.

The linear stability of a $q$-periodic orbit of $P$ such that
$P^q(z_0) = z_0$ is determined from the linearized-map matrix
$DP^q(z_0)$ of $P^q$ at an orbit point $z_0$.
Here $P^q$ means the $q$-times iterated map.
Using the Floquet theory, the matrix
$M$ $(\equiv DP^q)$ can be obtained by integrating the linearized
equations for small displacements,
\begin{mathletters}
\begin{eqnarray}
  \delta {\dot x} &=& \delta y,  \\
  \delta {\dot y} &=& - \Gamma \delta y + 2 \pi A \cos 2 \pi t
  \cos 2 \pi x \, \delta x
\end{eqnarray}
\label{eq:LE1}
\end{mathletters}
with two initial displacements $(\delta x, \delta y) = (1,0)$ and
$(0,1)$ over the period $q$. The eigenvalues, $\lambda_1$ and
$\lambda_2$, of $M$ are called the Floquet (stability) multipliers,
characterizing the orbit stability. By using the Liouville's
formula \cite{Arnold}, we obtain the determinant of $M$ $({\rm det}\;
M)$,
\begin{equation}
{\rm det}\;M= e^{-\Gamma q}.
\end{equation}
Hence the pair of Floquet multipliers of a periodic orbit lies either
on the circle  of radius  $e^{-\Gamma q/2}$ or on the  real axis
in the complex plane. The  periodic orbit  is stable when both
multipliers lie inside the unit circle. We first  note that they never
cross the unit circle, except at the real axis and hence Hopf
bifurcations do not occur. Consequently, it can lose its stability
only when a Floquet multiplier decreases (increases) through $-1$
$(1)$ on the real axis. When a Floquet multiplier $\lambda$ decreases
through $-1$, the periodic orbit loses its stability via
period-doubling bifurcation. On the other hand, when a Floquet
multiplier $\lambda$ increases through $1$, it becomes unstable via
pitchfork or saddle-node bifurcation. For each case of the
period-doubling and pitchfork bifurcations, two types of supercritical
and subcritical bifurcations occur. For more details on bifurcations,
refer to Ref.~\cite{Guckenheimer1}.

Lyapunov exponents of  an orbit $\{ z_m \}$  in the Poincar{\'{e}} map
$P$ characterize the mean exponential rate of divergence of nearby
orbits \cite{Lexp}.
There exist two Lyapunov exponents $\sigma_1$ and $\sigma_2$
($\sigma_1 \geq  \sigma_2$) such  that $\sigma_1  + \sigma_2 =  -
\Gamma$, because the linearized Poincar{\'{e}} map $DP$ has a constant
Jacobian determinant, det$DP = e^{-\Gamma}$.
We choose  an initial perturbation  $\delta z_0$ to  the initial orbit
point $z_0$ and iterate the linearized map $DP$ for
$\delta z$ along the orbit to obtain the magnitute $d_m$
$(\equiv |\delta z_m|)$ of $\delta z_m$.
Then, for  almost all infinitesimally-small  initial perturbations, we
have the largest Lyapunov exponent $\sigma_1$ given by
\begin{equation}
\sigma_1 = \lim_{m \rightarrow \infty}
   {1 \over m}  \ln {d_m \over d_0}.
\end{equation}
If $\sigma_1$ is positive, then the orbit is called a chaotic orbit;
otherwise, it is called a regular orbit.

\section{Rich dynamical behaviors of the damped MO}
\label{sec:DB}

In this section, by varying the amplitude $A$, we investigate the
evolutions of both the stationary points and the rotational orbits
of period $1$ in the damped MO for a moderately damped case of
$\Gamma=1.38$. As $A$ is increased, the damped MO, albeit simple
looking, exhibits rich dynamical behaviors, such as
symmetry-breaking pitchfork bifurcations, period-doubling transitions
to chaos, symmetry-restoring attractor-merging crises, and
saddle-node bifurcations giving rise to new periodic attractors.
In addition to these familiar behaviors,
the stationary points also undergo a cascade of resurrections (i.e.
an infinite sequence of alternating restabilizations and
destabilizations). It is found that the restabilizations
(destabilizations) occur via alternating subcritical (supercritical)
period-doubling and pitchfork bifurcations. An infinite sequence
of period-doubling bifurcations, leading to chaos, also follows
each destabilization of the stationary points.

\subsection{Evolution of the stationary points}
\label{subsec:SS}

 We first consider the case of the stationary points. The damped MO
has two stationary points $\hat{z}$'s. One is $\hat{z}_I = (0,0)$,
and the other one is $\hat{z}_{II} = ({1 \over 2},0)$. These
stationary points are symmetric ones with respect to the inversion
symmetry $S_2$, while they are asymmetric and conjugate ones with
respect to the shift symmetry $S_1$. Hence they are partially
symmetric orbits with only the inversion symmetry $S_2$. We also
note that the two stationary points are the fixed points of
the Poincar\'{e} map $P$ [i.e., $P(\hat{z})= \hat{z}\, (\hat{z} =
\hat{z}_I, \hat{z}_{II})$].

\begin{figure}
\epsfig{file={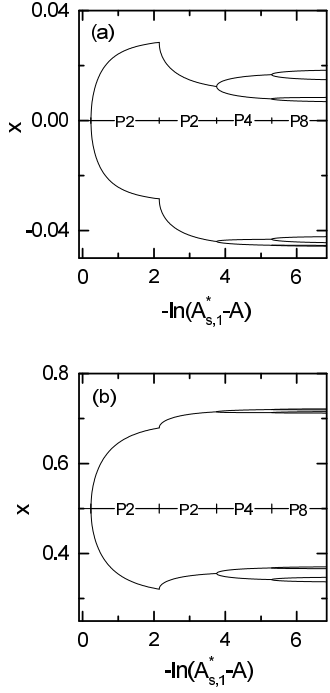}, width=\columnwidth} \vspace{-1.5cm}
\caption{Bifurcation diagrams starting from (a) the
         $S_2$-symmetric, but $S_1$-asymmetric, stationary point
         $\hat{z}_I$ and (b) its $S_1$-conjugate stationary point
         $\hat{z}_{II}$. The first and second P2's denote the stable
         $A$-ranges of the $S_2$-symmetric and $S_2$-asymmetric orbits
         of period 2, respectively. The other PN $(N=4,8)$ also
         designates the stable $A$-range of the $S_2$-asymmetric
         periodic orbit with period N.
         }
\label{fig:BD1}
\end{figure}

 With increasing $A$ we investigate the evolution of the two fixed
points $\hat{z}_I$ and $\hat{z}_{II}$. Two bifurcation diagrams
starting from $\hat{z}_I$ and $\hat{z}_{II}$ are given in
Figs.~\ref{fig:BD1}(a) and \ref{fig:BD1}(b),
respectively. Each fixed point loses its stability via
symmetry-conserving period-doubling bifurcation, giving rise to a
stable $S_2$-symmetric orbit with period $2$. However, as $A$ is
further increased each $S_2$-symmetric orbit of period $2$ becomes
unstable by a symmetry-breaking pitchfork bifurcation, leading to
the birth of a conjugate pair of $S_2$-asymmetric orbits of
period $2$. (For the sake of convenience, only one $S_2$-asymmetric
orbit of period $2$ is shown.) After breakdown of the $S_2$
symmetry, each $2$-periodic orbit with completely broken symmetries
exhibits an infinite sequence of period-doubling bifurcations,
ending at a finite critical point $A^*_{s,1}$
$(=3.934\,787\, \cdots)$. The critical scaling behaviors
near the critical point $A^*_{s,1}$ are the same as those for the
1D maps \cite{Feigenbaum}, as will be seen in Sec.~\ref{sec:CSB}.

\begin{figure}
 \epsfig{file={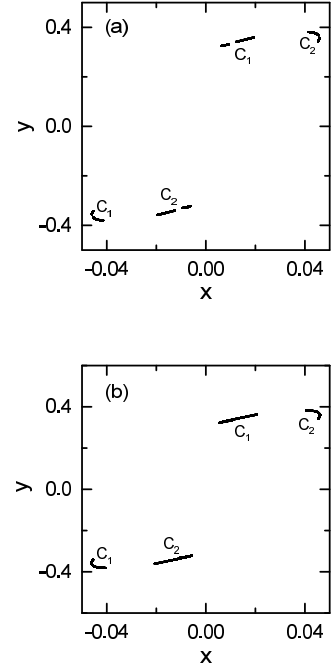}, width=\columnwidth}
\vspace{-1.5cm} \caption{Band-merging of chaotic attractors.
         (a) For $A=3.937$, each of the chaotic attractors $c_1$
         and $c_2$ is composed of four pieces. However, as $A$ is
         increased these pieces also merge to form two larger pieces.
         An example for $A=3.94$ is shown in (b).
         }
\label{fig:BM}
\end{figure}

 After the period-doubling transition to chaos, four small chaotic
attractors with completely broken symmetries appear; they are related
with respect to the two symmetries $S_1$ and $S_2$. As $A$ is
increased the different parts of each chaotic attractor coalease and
form larger pieces. For example, two chaotic attractors with
$\sigma_1$ (largest Lyapunov exponent) $\simeq 0.11$, denoted by $c_1$
and $c_2$, near the unstable stationary point $\hat{z}_I$ are shown
in Fig.~\ref{fig:BM}(a) for $A=3.937$; their conjugate chaotic
attractors with respect to the $S_1$ symmetry near the unstable
stationary point $\hat{z}_{II}$ are not shown. Each one is composed of
four distinct pieces. However, as $A$ is further increased these
pieces also merge into two larger pieces. An example with $\sigma_1
\simeq 0.18$ is shown in Fig.~\ref{fig:BM}(b) for $A=3.94$.

\begin{figure}
 \epsfig{file={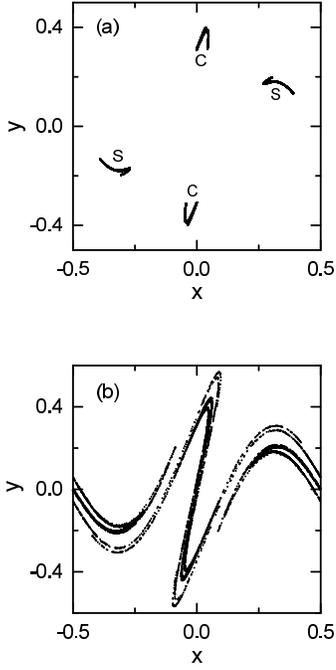}, width=\columnwidth}
\vspace{-1.5cm} \caption{Attractor-merging crises. The two chaotic
attractors $c_1$
         and $c_2$ in Fig.~2(b) merge into a larger one $c$ via
         $S_2$-symmetry-restoring crisis. For $A=3.96$ the chaotic
         attractor $c$ with the inversion symmetry $S_2$ and its
         conjugate one $s$ with respect to the $S_1$ symmetry are
         shown in (a). These two $S_1$-asymmetric small chaotic
         attractors $c$ and $s$ also merge to form a larger one
         via $S_1$-symmetry-restoring crisis. A single large chaotic
         attractor with completely restored $S_1$ and $S_2$ symmetries
         is shown in (b) for $A=3.98$.
         }
\label{fig:AM1}
\end{figure}

\begin{figure}
 \epsfig{file={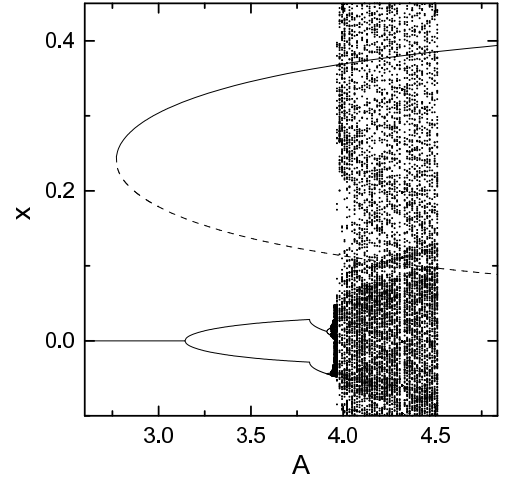}, width=\columnwidth}
\vspace{-3cm} \caption{Jump to a rotational orbit. The large
symmetric chaotic
         attractor in Fig.~3(b) disappears for $A \simeq 4.513$, and
         then the asymptotic state of the damped MO becomes a stable
         rotational orbit with period $1$ born via
         saddle-node bifurcation. Such a saddle-node bifurcation,
         giving rise to a pair of stable and unstable orbits of period
         $1$, occurs for $A \simeq 2.771$ (a stable one is denoted by a
         solid line, while an unstable one is represented by a dashed
         line). A bifurcation diagram starting from $\hat{z}_I$ is also
         shown.
         }
\label{fig:AD}
\end{figure}

As $A$ exceeds a critical value $(\simeq 3.9484)$, the two chaotic
attractors $c_1$ and $c_2$ in Fig.~\ref{fig:BM}(b) merge into a
larger one $c$ through an $S_2$-symmetry-restoring attractor-merging
crisis. For example, a chaotic attractor $c$ with $\sigma_1 \simeq
0.37$ and its conjugate one, denoted by $s$, with respect to the
$S_1$ symmetry are shown in Fig.~\ref{fig:AM1}(a) for $A=3.96$. These
two chaotic attractors $c$ and $s$ are $S_2$-symmetric ones, although
they are still $S_1$-asymmetric and conjugate ones. Thus the inversion
symmetry $S_2$ is first restored. However, as $A$ increases through a
second critical value $(\simeq 3.9672)$, the two small chaotic
attractors $c$ and $s$ also merge into a larger one via
$S_1$-symmetry-restoring attractor-merging crisis, as shown in
Fig.~\ref{fig:AM1}(b) for $A=3.98$. Note that the single large chaotic
attractor with $\sigma \simeq 0.64$ is both $S_1$- and $S_2$-symmetric
one. Consequently, the two symmetries $S_1$ and $S_2$ are completely
restored, one by one through two successive symmetry-restoring
attractor-merging crises. However, this large chaotic attractor
disappears for $A \simeq 4.513$, and then the system is asymptotically
attracted to a stable rotational orbit of period $1$ born through a
saddle-node bifurcation, as shown in Fig.~\ref{fig:AD}.

\subsection{Evolution of the rotational orbits}
\label{subsec:RS}

\begin{figure}
 \epsfig{file={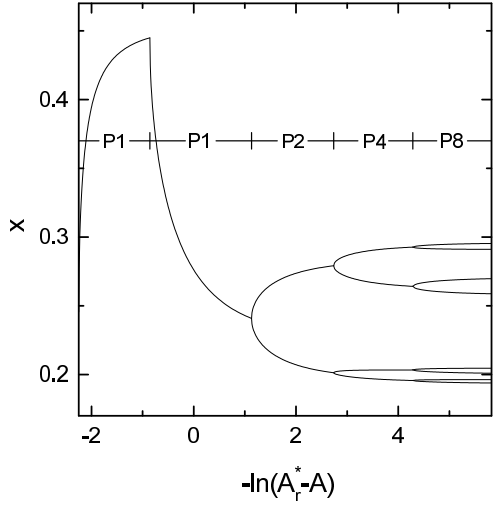}, width=\columnwidth}
\vspace{-3cm} \caption{Bifurcation diagram starting from the
$S_1$-symmetric, but
         $S_2$-asymmetric, rotational orbit with period $1$. The first
         and second P1's denote the stable $A$-ranges of the
         $S_1$-symmetric and $S_1$-asymmetric orbits of period 1,
         respectively. The other PN $(N=2,4,8)$ also designates the
         stable $A$-range of the $S_1$-asymmetric periodic orbit with
         period N.
         }
\label{fig:RBD}
\end{figure}

 We now investigate the evolution of the rotational orbits of
period $1$. A pair of stable and unstable rotational orbits with
period $1$ is born for $A \simeq 2.771$ via saddle-node bifurcation.
In contrast to the stationary points, these rotational orbits are
$S_1$-symmetric, but $S_2$-asymmetric and conjugate, ones.
The bifurcation diagram starting from a stable rotational orbit
with positive angular velocity is shown in Fig.~\ref{fig:RBD}.
(For convenience, the bifurcation diagram starting from its
$S_2$-conjugate rotational orbit with negative angular velocity is
omitted.) The $S_1$-symmetric rotational orbit of period $1$ becomes
unstable by a symmetry-breaking pitchfork bifurcation, which results
in the birth of a pair of $S_1$-asymmetric rotational orbits with
period $1$. (For the sake of conveninece, only one $S_1$-asymmetric
orbit of period $1$ is shown.) Then each rotational orbit with
completely broken symmetries undergoes an infinite sequence of
period-doubling bifurcations, accumulating at a finite critical
point $A^*_r$ $(=12.252\,903\, \cdots)$. The critical behaviors
near the accumulation point $A^*_r$ are also the same as those for
the 1D maps, as in the case of the stationary points.

\begin{figure}
 \epsfig{file={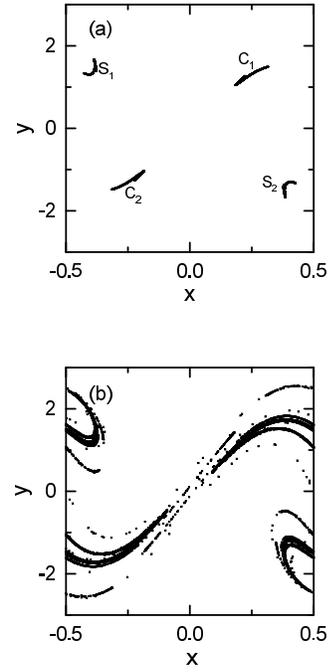}, width=\columnwidth}
\vspace{-1.5cm} \caption{Attractor-merging crisis for the
rotational case. Through a
         band-merging process, each of the four chaotic attractors
         $c_1$, $c_2$, $s_1$, and $s_2$ with completely broken
         symmetries eventually becomes composed of a single piece, as
         shown in (a) for $A=12.32$.
         The four small asymmetric chaotic attractors merge into a
         larger one via symmetry-restoring crisis. A single large
         chaotic attractor with simultaneously restored $S_1$ and
         $S_2$ symmetries is shown in (b) for $A=12.38$.
     }
\label{fig:AM2}
\end{figure}

For $A> A^*_r$, four chaotic attractors with completely broken
symmetries appear; they are related with respect to the two
symmetries $S_1$ and $S_2$. Through a band-merging process, each
chaotic attractor eventually becomes composed of a single piece, as
shown in Fig.~\ref{fig:AM2}(a) for $A=12.32$. Four chaotic attractors
with $\sigma_1 \simeq 0.36$ are denoted by $c_1$, $c_2$, $s_1$, and
$s_2$, respectively. However, as $A$ passes through a critical value
$(\simeq 12.3424)$ the four small chaotic attractors merge into a larger
one via symmetry-restoring attractor-merging
crisis. An example for $A=12.38$ is given in Fig.~\ref{fig:AM2}(b).
Note that the single large chaotic attractor with
$\sigma_1 \simeq 0.64$ has both the $S_1$ and $S_2$ symmetries.
Thus the two symmetries are completely restored through one
symmetry-restoring attractor-merging crisis, which is in
contrast to the case of the stationary points.

However, as shown in Fig.~\ref{fig:IM1}, the large symmetric chaotic
attractor in Fig.~\ref{fig:AM2}(b) also disappears for $A \simeq
13.723$, at which saddle-node bifurcations occur. After disappearence
of this large chaotic attractor, the system is asymptotically
attracted to a stable $S_2$-symmetric, but $S_1$-asymmetric, orbit of
period $2$ born via saddle-node bifurcation. This stable
$S_2$-symmetric orbit with period $2$ also exhibits rich dynamical
behaviors similar to those of the stationary points. That is, as $A$
is increased, a symmetry-breaking pitchfork bifurcation,
period-doubling transition to chaos, merging of small asymmetric
chaotic attractors into a large symmetric one via symmetry-restoring
attractor-merging crisis, and so on are found. However, unlike the
cases of the stationary points and the rotational orbits, the large
symmetric chaotic attractor disappears for $A=23.751\,799\, \cdots$,
at which the two unstable stationary points $\hat{z}_I$ and
$\hat{z}_{II}$ become restabilized through subcritical
period-doubling bifurcations. These ``resurrections'' of the
stationary points will be described below in some details.

\subsection{Resurrections of the stationary points}
\label{subsec:RSS}

The linear stability of the two stationary points
$\hat{z}_I$ and $\hat{z}_{II}$ is determined by their linearized
equations,
\begin{equation}
 {\delta {\ddot x}} + \Gamma \, {\delta {\dot x}} \mp
 2 \pi A \cos 2 \pi t \, {\delta x} =0,
\label{eq:LE2}
\end{equation}
where $-$ $(+)$ sign of the third term corresponds to the case of
${\hat z}_I$ (${\hat z}_{II}$). (The linearized equation of
${\hat z}_{II}$ can also be transformed into that of
${\hat z}_I$ by just making a shift in time, $t \rightarrow t +
{1 \over 2}$.) Note that Eq.~(\ref{eq:LE2}) is just a simple
form of the more general damped Mathieu equation \cite{Mathieu}.
It is well known that the Mathieu
equation has an infinity of alternating stable and unstable
$A$ ranges. Hence, as $A$ is increased, the stationary
points undergoes a cascade of ``resurrections,'' i.e., they will
restabilize after they lose their stability, destabilize again, and
so forth {\it ad infinitum}.

\begin{figure}
 \epsfig{file={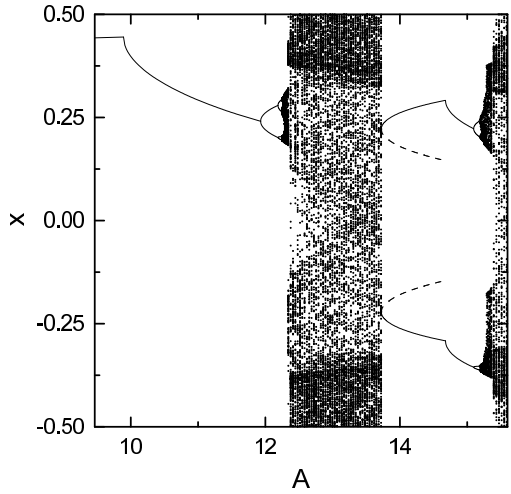}, width=\columnwidth}
\vspace{-3cm} \caption{Jump to an orbit with period $2$. The large
symmetric chaotic
         attractor in Fig.~6(b) disappears for $A \simeq 13.723$, at
         which a saddle-node bifurcation, giving rise to a pair of
         stable and unstable orbits with period $2$, occurs. (A stable
         one is denoted by a solid line, while an unstable one is
         represented by a dashed line). Then the damped MO is
         asymptotically attracted to the stable $S_2$-symmetric,
         but $S_1$-asymmetric, orbit with period $2$. This stable
         $S_2$-symmetric orbit also exhibits rich dynamical behaviors
         similar to those of the stationary points. In the left part,
         a bifurcation diagram starting from a rotational orbit with
         period $1$ is also given.
         }
\label{fig:IM1}
\end{figure}

\begin{figure}
 \epsfig{file={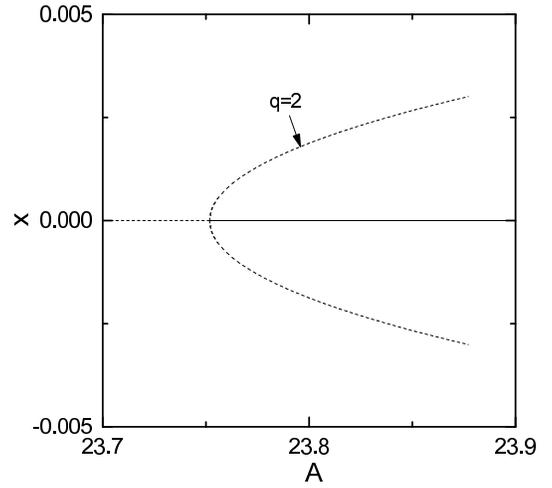}, width=\columnwidth}
\vspace{-3cm} \caption{First resurrection of the stationary point
$\hat{z}_I$.
         When $A$ passes through the first restabilization value
         $(\simeq 23.752)$, the unstable stationary point
         $\hat z_I$ restabilizes through a subcritical
         period-doubling bifurcation, giving rise to an unstable
         orbit with period $q=2$. The solid and dashed
         lines also denote stable and unstable orbits, respectively.
     }
\label{fig:RS1}
\end{figure}

\begin{figure}
 \epsfig{file={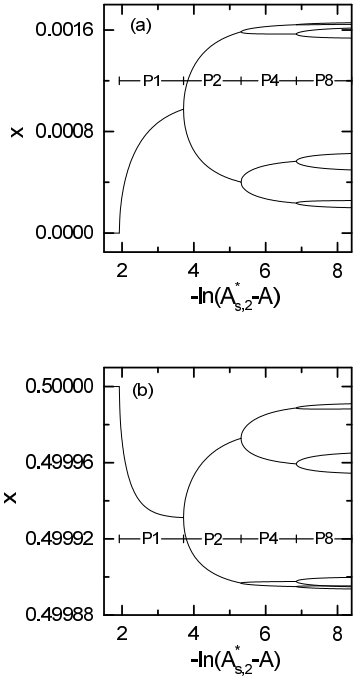}, width=\columnwidth}
\vspace{-1.5cm} \caption{Seoncd bifurcation diagrams starting from
(a) the
         restabilized stationary point $\hat{z}_I$ and (b) its
         $S_1$-conjugate stationary point $\hat{z}_{II}$.
         Here the PN designates the stable $A$-range of the
         $S_2$-asymmetric periodic orbit with period N (N=$1,2,4,8$).
     }
\label{fig:BD2}
\end{figure}

It is found that their restabilizations (destabilizations) occur
through alternating subcritical (supercritical) period-doubling
and pitchfork bifurcations. As examples, we consider
the first and second resurrections of the stationary points.
The first resurrection of $\hat{z}_I$ is shown in Fig.~\ref{fig:RS1}.
When $A$ passes through the first restabilization value
$(=23.751\,799\, \cdots)$, the rightmost large symmetric chaotic
attractor in Fig.~\ref{fig:IM1} disappears and the unstable
stationary point $\hat{z}_I$ restabilizes via subcritical
period-doubling bifurcation, giving rise to an
unstable orbit of period $2$. Two bifurcation diagrams starting from
the restabilized $\hat{z}_I$ and $\hat{z}_{II}$ are given in
Fig.~\ref{fig:BD2}(a) and \ref{fig:BD2}(b), respectively. Each
stationary point loses its stability via symmetry-breaking pitchfork
bifurcation, giving rise to a pair of $S_2$-asymmetric orbits with
period $1$; only one asymmetric $1$-periodic orbit is shown. This is
in contrast to the case given in Sec.~\ref{subsec:SS} (compare
Fig.~\ref{fig:BD1} with Fig.~\ref{fig:BD2}), where the stationary
points become unstable via symmetry-conserving period-doubling
bifurcations. After breakdown of the $S_2$ symmetry, an infinite
sequence of period-doubling bifurcations follows and ends at its
accumulation point $A^*_{s,2}$
$(=24.148\,001\, \cdots)$. When $A$ exceeds $A^*_{s,2}$, a second
period-doubling transition to chaos occurs. The critical scaling
behaviors of period doublings near the second critical point
$A=A^*_{s,2}$ are also the same as those near the first critical
point $A^*_{s,1}$.

\begin{figure}
 \epsfig{file={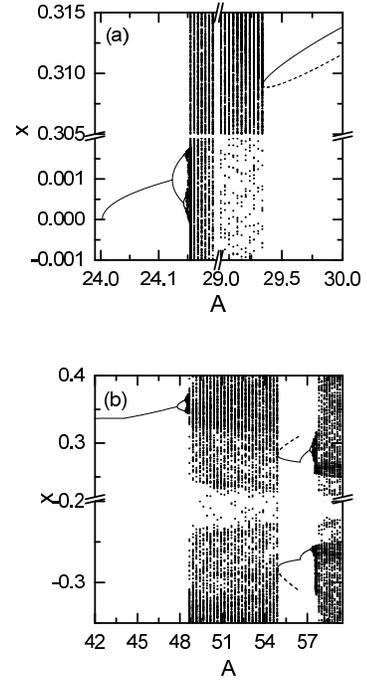}, width=\columnwidth}
\vspace{-1.5cm} \caption{Dynamical behaviors after the second
period-doubling
         transition to chaos is shown in (a). For $A \simeq 24.1549$
         small chaotic attractors with completely broken symmetries
         merge into a large symmetric chaotic attractor via
         symmetry-restoring crisis. However, this large symmetric
         chaotic attractor disappears for $A \simeq 29.342$, and then
         the damped MO is asymptotically attracted to a stable orbit
         of period $1$ born via saddle-node bifurcation. As shown in
         (b), subsequent evolution of the stable $S_1$-symmetric, but
         $S_2$-asymmetric, $1$-periodic orbit is similar to that of
         the rotational orbit shown in Fig.~7. Here the solid and
         dashed lines also denote stable and unstable orbits,
         respectively. For other details see text.
     }
\label{fig:IM2}
\end{figure}

Dynamical behaviors of the damped MO after the second period-doubling
transition to chaos are shown in Fig.~\ref{fig:IM2}(a). As $A$ passes
through a critical value $(\simeq 24.1549)$, small chaotic attractors
with completely broken symmetries merge into a large symmetric chaotic
attractor via symmetry-restoring attractor-merging crisis. However,
the large chaotic attractor also disappears for $A \simeq 29.342$, at
which saddle-node bifurcations occur. After disappearence of the
large chaotic attractor, the damped MO is asymptotically attracted to
a stable $S_1$-symmetric, but $S_2$-asymmetric, orbit of period $1$
born via saddle-node bifrucation. The subsequent evolution of the
stable $1$-periodic orbit is shown in Fig.~\ref{fig:IM2}(b). Note that
it is similar to that of the rotational orbit described in IIIB
(compare Fig.~\ref{fig:IM2}(b) with Fig.~\ref{fig:IM1}).

\begin{figure}
 \epsfig{file={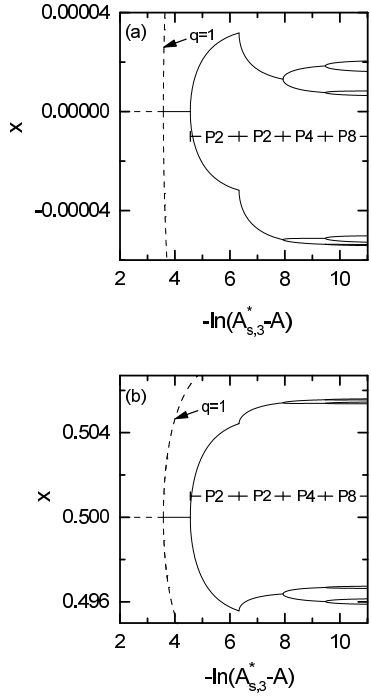}, width=\columnwidth}
\vspace{-1.5cm} \caption{Second resurrections of the stationary
points and third
         bifurcation diagrams starting from (a) the restabilized
         stationary point $\hat{z}_I$ and (b) its $S_1$-conjugate
         stationary point $\hat{z}_{II}$. When $A$ passes through the
         second restabiliztion value $(\simeq 67.08)$, each unstable
         stationary point becomes restabilized via subcritical
         pitchfork bifurcation, giving rise to a pair of unstable
         orbits with period $q=1$. The solid and dashed lines
         represent stable and unstable orbits, respectively.
         The third bifurcation diagrams are similar to those in
         Fig.~1; the symbols are also the same as those of Fig.~1.
     }
\label{fig:BD3}
\end{figure}

The rightmost large symmetric chaotic attractor in
Fig.~\ref{fig:IM2}(b) appears via symmetry-restoring attractor-merging
crisis for $A=57.67$. However, it also disappears for
$A=67.076\,913\, \cdots$, at which each stationary point restabilizes
again. Unlike the case of the first resurrection (see
Fig.~\ref{fig:RS1}), this second resurrection of each stationary
point occurs via subcritical pitchfork bifurcation, giving rise to a
pair of unstable $1$-periodic orbits with broken symmetries. The
second resurrections of the two stationary points $\hat{z}_I$ and
$\hat{z}_{II}$ and their subsequent bifurcation diagrams are shown in
Fig.~\ref{fig:BD3}(a) and \ref{fig:BD3}(b), respectively. Note that
these third bifurcation diagrams are similar to those in
Fig.~\ref{fig:BD1}. The critical scaling behaviors near the third
period-doubling transition point $A^*_{s,3}$ $(=67.104\,872\, \cdots)$
are also the same as those near the first period-doubling
transition point $A=A^*_{s,1}$.

\section{Critical scaling behaviors in the period-doubling cascades}
\label{sec:CSB}

In this section, we study the critical scaling behaviors in the
period-doubling cascades. The orbital scaling behavior and
the power spectra of the periodic orbits born via period-doubling
bifurcations as well as the parameter scaling behavior are
particularly investigated.

The critical scaling behaviors for all cases studied are found to be
the same as those for the 1D maps. As an example, we consider the
first period-doubling transition to chaos for the case of the
stationary points. As explained in Sec.~\ref{subsec:SS},
each stationary point becomes unstable through symmetry-conserving
period-doubling bifurcation, giving rise to a stable $S_2$-symmetric
orbit of period $2$. However, each $S_2$-symmetric orbit of period
$2$ also becomes unstable via symmetry-breaking pitchfork
bifurcation, which results in the birth of a conjugate pair of
$S_2$-asymmetric orbits with period $2$. Then, each $2$-periodic
orbit with completely broken symmetries undergoes an
infinite sequence of period-doubling bifurcations, ending at its
accumulation point $A^*_{s,1}$. Table \ref{tab:PSB} gives the $A$
values at which the period-doubling bifurcations occur; at $A_k$,
a Floquet multiplier of an asymmetric orbit with period $2^k$
becomes $-1$. The sequence of
$A_k$ converges geometrically to its limit value $A^*_{s,1}$ with
an asymptotic ratio $\delta$:
\begin{equation}
\delta_k = {{A_k - A_{k-1}} \over {A_{k+1} - A_k}} \rightarrow \delta.
\end{equation}
The sequence  of $\delta_k$  is also  listed in Table I. Note
that its limit   value  $\delta$   $(\simeq   4.67)$  agrees  well
with that $(=4.669\cdots)$ for the 1D maps \cite{Feigenbaum}.
We also obtain the value of $A^*_{s,1}$ $(=3.934\,787\,024)$
by superconverging the sequence of $\{ A_k \}$ \cite{MacKay}.

As in the 1D maps, we are interested in the orbital scaling behavior
near the most rarified region. Hence, we first locate
the most rarified region by choosing an orbit point $z^{(k)}$
$[=(x^{(k)},y^{(k)})]$ which has the largest distance from its
nearest orbit point $P^{2^{k-1}}(z^{(k)})$ for
$A=A_k$. The two sequences $\{ x^{(k)} \}$ and  $\{ y^{(k)} \}$,
listed in Table \ref{tab:OS}, converge geometrically to their
limit values $x^*$ and $y^*$ with the 1D asymptotic ratio
$\alpha$ $(=-2.502\, \cdots)$, respectively:
\begin{equation}
\alpha_{x,k} = { {x^{(k)} - x^{(k-1)}} \over {x^{(k+1)} - x^{(k)}} }
\rightarrow \alpha, \;\;
\alpha_{y,k} = {{y^{(k)} - y^{(k-1)}} \over
{y^{(k+1)} - y^{(k)}}} \rightarrow \alpha.
\end{equation}
The values of $x^*$ $(=0.015\,369)$ and $y^*$ $(=0.348\,175)$ are also
obtained by superconverging the sequences of $x^{(k)}$ and $y^{(k)}$,
respectively.

\begin{figure}
 \epsfig{file={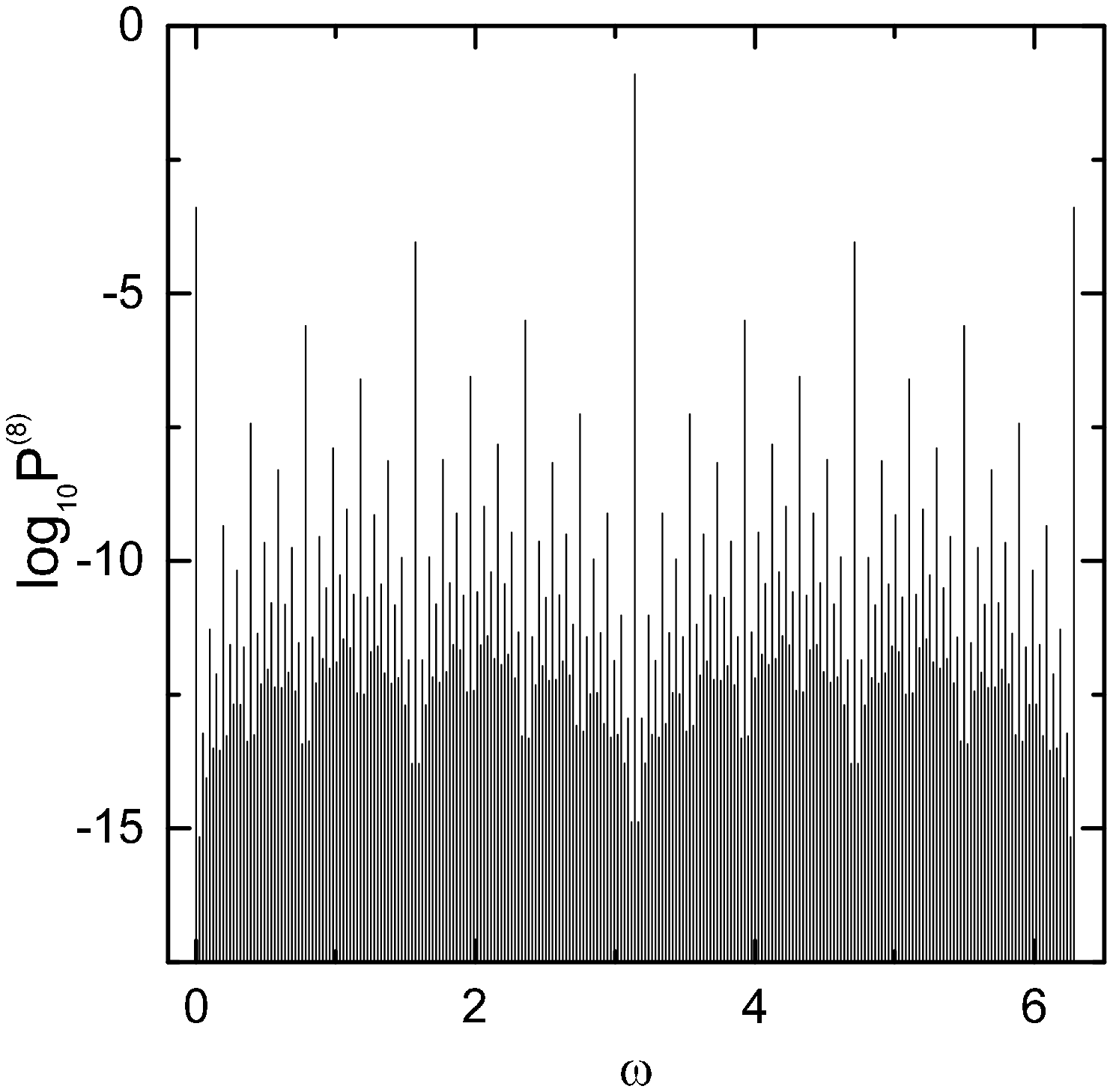}, width=\columnwidth}
\vspace{-3cm} \caption{Power spectrum $P^{(8)}(\omega)$ of level
$8$ for
          $A=A_8$ $(=3.934\,786\,542\,923).$
     }
\label{fig:PS}
\end{figure}

We also study the power spectra of the $2^k$-periodic orbits
at the period-doubling bifurcation points $A_k$.
Consider the orbit of level $k$ whose period is $q=2^k$,
$\{ z^{(k)}_m=(x^{(k)}_m,y^{(k)}_m), \; m=0,1,\ldots,q-1 \}$. Then
its Fourier component of this $2^k$-periodic orbit is given by
\begin{equation}
z^{(k)}(\omega_j) =   {1  \over q}  \sum_{m=0}^{q-1}   z^{(k)}_m e^{-i
\omega_j m},
\end{equation}
where $\omega_j = 2 \pi j / q$, and $j=0,1,\ldots,q-1$.
The power spectrum $P^{(k)}(\omega_j)$ of level $k$ defined by
\begin{equation}
P^{(k)}(\omega_j) = |z^{(k)}(\omega_j)|^2,
\end{equation}
has discrete peaks at $\omega = \omega_j$.
In the  power spectrum  of the  next $(k+1)$ level,  new peaks  of the
$(k+1)$th generation appear at odd harmonics of the fundamental
frequency, $\omega_j = 2 \pi (2j+1) / 2^{(k+1)}$ $(j=0, \ldots,
2^k -1)$. To  classify  the  contributions  of  successive
period-doubling bifurcations  in  the  power spectrum of level $k$, we
write
\begin{equation}
P^{(k)}= P_{00} \delta(\omega) +  \sum_{l=1}^{k}
\sum_{j=0}^{2^{(l-1)}-1}
P^{(k)}_{lj} \delta(\omega- \omega_{lj}),
\end{equation}
where  $P^{(k)}_{lj}$  is   the  height  of  the $j$th   peak  of  the
$l$th generation
appearing at $\omega=\omega_{lj}$ $(\equiv 2 \pi (2j+1) / 2^l)$.
As an example, we consider the power spectrum $P^{(8)}(\omega)$ of
level $8$ shown in Fig.~\ref{fig:PS}. The average height of the peaks
of the $l$th generation is given by
\begin{equation}
\phi^{(k)}(l)   =    {1   \over    2^{(l-1)}}   \sum_{j=0}^{2^{l-1}-1}
P_{lj}^{(k)}.
\end{equation}
It is of interest whether or not the sequence of the ratios of the
successive average heights
\begin{equation}
 2 \beta^{(k)}(l) = \phi^{(k)}(l) / \phi^{(k)}(l+1),
\end{equation}
converges. The  ratios are  listed  in  Table \ref{tab:PSS}. They
seem to approach a limit value, $2 \beta \simeq 21$, which also agrees
well with that $(=20.96 \cdots)$ for the 1D maps \cite{Rudnick}.

\section{Summary}

Dynamical behaviors of the damped MO are investigated in details by
varying the amplitude $A$. The damped MO, despite its apparent
simplicity, exhibits rich dynamical behaviors such as breakdown of
symmetries via pitchfork bifurcations, period-doubling transitions
to chaos, restoration of symmetries via attractor-merging
crises, and creation of new periodic attractors through saddle-node
bifurcations. In addition to these familiar behaviors, a cascade of
resurrections (i.e., an infinite sequence of alternating
restabilizations and destabilizations) of the stationary points
occurs. It is found that the stationary points restabilize
(destabilize) via alternating subcritical (supercritical)
period-doubling and pitchfork bifurcations. An infinite sequence of
period-doubling bifurcations, leading to chaos, also follows each
destabilization of the stationary points. The critical scaling
behaviors in the period-doubling cascades are found to be the same
as those of the 1D maps.

\label{sec:SUM}

\acknowledgements
The author would like to thank K. Lee for his assitance in the
numerical computations. This work was supported by the Biomedlab Inc.
and by the Korea Research Foundation under Project No. 1998-015-D00065.

\begin{table}
\caption{ Asymptotically geometric convergence of the parameter
          sequence $\{ A_k \}$
        }
\label{tab:PSB}
\begin{tabular}{ccc}
$k$ & $A_k$ & $\delta_k$  \\
\tableline
1 & 3.911\,404\,100\,371 &        \\
2 & 3.929\,795\,227\,873 &   4.690 \\
3 & 3.933\,716\,964\,019 &   4.664 \\
4 & 3.934\,557\,747\,089 &   4.667 \\
5 & 3.934\,737\,918\,915&   4.669\\
6 & 3.934\,776\,506\,700&   4.668 \\
7 & 3.934\,784\,773\,689&   4.673 \\
8 & 3.934\,786\,542\,923&
\end{tabular}
\end{table}

\begin{table}
\caption{ Asymptotically  geometric   convergence  of   the  orbital
           sequences
          $\{ x^{(k)} \}$ and $\{ y^{(k)} \}$.
        }
\label{tab:OS}
\begin{tabular}{cccccc}
$k$ & $x^{(k)}$ & $\alpha_{x,k}$ & $y^{(k)}$ & $\alpha_{y,k}$   \\
\tableline
1 & 0.012\,394\,993 &          &  0.337\,125\,704 &         \\
2 & 0.016\,703\,927 &   -2.354 &  0.350\,786\,135 &  -3.410 \\
3 & 0.014\,873\,276 &   -2.607 &  0.346\,780\,101 &  -2.133 \\
4 & 0.015\,575\,414 &   -2.451 &  0.348\,658\,607 &  -2.714 \\
5 & 0.015\,288\,893 &   -2.532 &  0.347\,966\,417 &  -2.395 \\
6 & 0.015\,402\,057 &   -2.487 &  0.348\,255\,392 &  -2.561 \\
7 & 0.015\,356\,562 &   -2.511 &  0.348\,142\,575 &  -2.472 \\
8 & 0.015\,374\,678 &          &  0.348\,188\,212 &
\end{tabular}
\end{table}

\begin{table}
\caption{ Sequence $2\beta^{(k)}(l)$
$[\equiv \phi^{(k)}(l)/\phi^{(k)}(l+1)]$
of the ratios of the successive average heights of the peaks in the
power spectra
        }
\label{tab:PSS}
\begin{tabular}{cccccc}
\multicolumn{1}{c}{$k$} & \multicolumn{5}{c}{$l$} \\
 & $3$ & $4$ & $5$ & $6$ & $7$  \\
\tableline
6 & 18.2 & 22.5 & 21.0 & & \\
7 & 18.1 & 22.1 & 21.1 & 21.5 & \\
8 & 18.1 & 22.0 & 20.7 & 21.6 & 21.4
\end{tabular}
\end{table}

\end{multicols}


\begin{references}

\bibitem{Croquette} V. Croquette and C. Poitou, J. Phys.\ Lett.\
  {\bf 42}, 537 (1981).
\bibitem{Schmidt1} H. Meissner and G. Schmidt, Am. J. Phys.\ {\bf 54},
  800 (1986).
\bibitem{Briggs} K. Briggs, Am.\ J. Phys.\ {\bf 55}, 1083 (1987).
\bibitem{Bialek} J. Bialek, G. Schmidt, and B. H. Wang, Physica D
 {\bf 14}, 265 (1985).
\bibitem{Schmidt2} G. Schmidt, Comments Plasma Phys. Controlled
 Fusion {\bf 7}, 87 (1982).
\bibitem{Escande} D. F. Escande and F. Doveil, Phys.\ Lett.\ A
{\bf 83}, 307 (1981); J. Stat.\ Phys.\ {\bf 26}, 257 (1981).
\bibitem{Arnold} V. I. Arnold, {\it Ordinary Differential Equations}
(MIT Press, Cambridge, 1973), p.\ 114.
\bibitem{Guckenheimer1} J. Guckenheimer and P. Holmes, {\it Nonlinear
 Oscillations, Dynamical Systems, and Bifurcations of Vector Fields}
  (Springer-Verlag, New York, 1983), Sec.~3.5.
\bibitem{Feigenbaum} M. J. Feigenbaum, J. Stat.\ Phys.\ {\bf 19}, 25
 (1978); {\bf 21}, 669 (1979).
\bibitem{Crisis} C. Grebogi, E. Ott, F. Romeiras, and J. A. Yorke,
 Phys.\ Rev.\ A {\bf 36}, 5365 (1987)
\bibitem{Kim} S.-Y. Kim and K. Lee, Phys. Rev. E {\bf 53}, 1579
 (1996).
\bibitem{Guckenheimer2} J. Guckenheimer and P. Holmes, {\it Nonlinear
 Oscillations, Dynamical Systems, and Bifurcations of Vector Fields}
  (Springer-Verlag, New York, 1983), p.\ 24.
\bibitem{Lexp} A. J. Lichtenberg and M. A. Lieberman, {\it Regular and
 Chaotic Dynamics} (Springer-Verlag, New York, 1983), p.\ 262.
\bibitem{Mathieu} P. M. Morse and H. Feshbach, {\it Methods of
 Theoretical Physics} (McGraw-Hill, New York, 1953), Sec.~5.2;
 J. Mathews and R. L. Walker, {\it Mathematical Methods of Physics}
 (Benjamin, New York, 1965), Sec.~7.5.
\bibitem{MacKay} R. S. MacKay, Ph.D. thesis, Princeton University, 1982.
 See Eqs.~3.1.2.12 and 3.1.2.13.
\bibitem{Rudnick} M. Nauenberg and J. Rudnick, Phys.\ Rev.\ B 24, 493
 (1981).

\end{references}
\end{document}